\documentclass[aps,prb,twocolumn,nofootinbib,showpacs,preprintnumbers,notitlepage,superscriptaddress]{revtex4-2}

\usepackage{graphicx}
\usepackage{graphics}
\usepackage{amsmath}
\usepackage{amssymb}
\usepackage{amsfonts}
\usepackage{dsfont}
\usepackage{braket}
\usepackage{color}
\usepackage{braket,slashed}
\usepackage[mathscr]{euscript}
\definecolor{darkblue}{rgb}{0, 0, 0.8}
\usepackage[colorlinks=true, breaklinks=true, linkcolor=red, citecolor=blue, urlcolor=blue]{hyperref} 
\usepackage{subfigure}
\usepackage{xfrac}
\usepackage{bm}
\usepackage{kantlipsum}
\usepackage{enumitem}
\usepackage{tikz}
\usepackage{framed}
\usepackage{graphicx}
\usepackage{subfigure}
\usepackage{cleveref}
\usepackage{adjustbox}

\usepackage{xcolor}

\allowdisplaybreaks[1]

\newcommand{\code}[1]{\texttt{#1}}

\newcommand{\e}{\ensuremath{\mathrm{e}}}

\newcommand{\He}{\ensuremath{H_{\mathrm{ent}}}}
\newcommand{\vS}{\ensuremath{\vec{S}}}

\newcommand{\diagram}[1]{\;\vcenter{\hbox{\includegraphics[scale=0.32,page=#1]{./diagrams.pdf}}}\;}
\newcommand{\tmdiags}[2]{\;\vcenter{\hbox{\includegraphics[scale=#2,page=#1]{./tmdiags.pdf}}}\;}

\begin{document}

\title{Simulating Chiral Ppin Liquids with Projected Entangled-Pair States}

\author{Juraj Hasik}
\email{j.hasik@uva.nl}
\affiliation{Laboratoire de Physique Th\'{e}orique, C.N.R.S. and Universit\'{e} de Toulouse, 31062 Toulouse, France}
\affiliation{Institute for Theoretical Physics, University of Amsterdam, Science Park 904, 1098 XH Amsterdam, The Netherlands}

\author{Maarten Van Damme}
\affiliation{Department of Physics and Astronomy, University of Ghent, Krijgslaan 281, 9000 Gent, Belgium}

\author{Didier Poilblanc}
\affiliation{Laboratoire de Physique Th\'{e}orique, C.N.R.S. and Universit\'{e} de Toulouse, 31062 Toulouse, France}

\author{Laurens Vanderstraeten}
\email{laurens.vanderstraeten@ugent.be}
\affiliation{Department of Physics and Astronomy, University of Ghent, Krijgslaan 281, 9000 Gent, Belgium}

\begin{abstract}
Doubts have been raised on the representation of chiral spin liquids  exhibiting topological order in terms of projected entangled pair states (PEPSs).  Here, starting from a simple spin-1/2 chiral frustrated Heisenberg model, we show that a faithful representation of the chiral spin liquid phase is in fact possible in terms of a generic PEPS upon variational optimization. We find a perfectly chiral gapless edge mode and a rapid decay of correlation functions at short distances consistent with a bulk gap, concomitant with a gossamer long-range tail originating from a PEPS bulk-edge correspondence. For increasing bond dimension, (i) the rapid decrease of spurious features -- SU(2) symmetry breaking and long-range tails in correlations -- together with (ii) a faster convergence of the ground state energy as compared to state-of-the-art cylinder matrix-product state simulations involving far more variational parameters, prove the fundamental relevance of the PEPS ansatz for simulating systems with chiral topological order. 
\end{abstract}

\maketitle

%%%%%%%%%%%%%%%%%%
\par\noindent\emph{\textbf{Introduction---}} %
%%%%%%%%%%%%%%%%%%
A chiral spin liquid (CSL) is a strongly correlated two-dimensional quantum spin system that serves as the lattice analog of a fractional quantum Hall system \cite{Kalmeyer1987, Kalmeyer1989, Wen1989}. It hosts a disordered ground state, and inherits many of the characteristic features of fractional quantum Hall physics, such as fractionalized excitations and protected chiral edge modes. The CSL was originally proposed on the level of a trial wave function, and later parent Hamiltonians were found \cite{Schroeter2007, Thomale2009, Nielsen2013} that stabilize a CSL ground state. Although these are not expected to describe real materials, in recent years many simpler model Hamiltonians were found that seem to realize CSL physics \cite{Messio2012, Bauer2014, Gong2014, He2014, Hu2015, Gong2015, Wietek2015, Wietek2017, Hickey2017, Gong2017, Motruk2020, Cookmeyer2021}. In these works, the use of performant numerical techniques such as variational matrix-product state (MPS) methods on cylindrical geometries, exact diagonalization on finite clusters or variational Monte Carlo approaches was crucial for finding evidence of CSL features. Accurate numerical simulations are instrumental in relating theoretical descriptions of CSL physics to real materials that can be probed in experiments; for example, a few materials were found that realize the triangular-lattice Hubbard model \cite{Shimizu2003, Itou2007, Law2017, Ni2019, Miksch2021, Ruan2021}, which possibly exhibits an emergent CSL phase between the metallic and the ordered insulating phases \cite{Szasz2020, BBChen2021, Tocchio2021}.

%\begin{figure}
%    \centering
%    \includegraphics[width=\columnwidth]{./cartoon.pdf}
%    \caption{\textbf{Cartoon picture of a chiral PEPS.} The virtual bonds between the sites are comprised of exponentially decaying correlations (black lines) and algebraically decaying correlations (light blue lines). Physical operators only couple to the exponential part, hence generating a bulk gap, whereas the algebraic part generates the long-range boundary hamiltonian.}
%    \label{fig:peps}
%\end{figure}

\par In recent years, projected entangled-pair states (PEPSs) \cite{Verstraete2004} have demonstrated a remarkable ability to simulate the ground states of strongly correlated lattice models in two dimensions~\cite{corboz2013,corboz2014,zheng2017,Hasik2021,Chen2021}. This class of tensor-network states can be formulated \cite{Jordan2008} and variationally optimized \cite{Corboz2016, Vanderstraeten2016} directly in the thermodynamic limit. The control parameters are the bond dimension $D$, controlling the amount of entanglement in the PEPS wave function, and the environment bond dimension $\chi$, which is introduced when contracting the PEPS approximately with boundary MPS \cite{Verstraete2004, Fishman2018, Vanderstraeten2022} or the corner-transfer matrix renormalization group (CTMRG) \cite{Baxter1968, Baxter1978, Nishino1996, Nishino1997, Orus2009, Corboz2010}. When simulating a critical model, a variational PEPS wave function typically exhibits a finite correlation length, which provides a length scale that can be used to extrapolate variational PEPS results \cite{Corboz2018, Rader2018, Czarnik2019, Vanhecke2021}. In the case of nonchiral spin liquids, this procedure has proven very powerful \cite{Hasik2021}.

\par The case of \emph{chiral} systems is less straightforward: Whereas nonchiral topological order can be described by PEPSs in a natural way \cite{Schuch2010, Schuch2011, Sahinouglu2021} -- for example, the resonating valence bond state on the kagome lattice is a simple example of gapped $\mathbb{Z}_2$ spin liquid and is written as a PEPS with bond dimension $D=3$ \cite{Poilblanc2012} -- capturing a gapped chiral topological wave function seems to be much harder. The Gaussian fermionic PEPS wave functions that were proposed exhibit algebraic decay of correlations \cite{Wahl2013, Wahl2014, Mortier2020}, a feature that was \emph{proven} necessary for representing chiral free-fermionic states as PEPSs \cite{Dubail2015}, see also \cite{Kapustin2020}. By Gutzwiller projecting two of these Gaussian PEPSs, a CSL wave function can be constructed, exhibiting an entanglement spectrum with the degeneracy pattern of a chiral conformal field theory. Subsequently, a class of SU(2) invariant PEPSs was proposed that exhibits chiral entanglement spectra \cite{Poilblanc2015, Poilblanc2016, Hackenbroich2018}. Motivated by the properties of these trial wavefunctions, microscopic CSL Hamiltonians were successfully studied with PEPS, where both SU(2) and spatial symmetries were imposed on the PEPS tensors \cite{Poilblanc2017, Chen2018}; recently, this approach was even extended to study SU($N$) CSLs \cite{Chen2020, Chen2021}. In all these cases, however, the PEPS exhibits a decay of correlations slower than exponential, i.e. an \emph{infinite} correlation length, which indicates an obstruction for simulating gapped chiral topological phases with PEPSs, in line with the no-go theorem for the free-fermionic case. This, in turn, makes these types of simulations very challenging numerically, all of which has led to the view that PEPSs are ill-suited for simulating chiral models.

\par In this paper, we want to disprove this view by performing an \textit{unconstrained} variational PEPS simulation of an SU(2) CSL model on the square lattice. Comparing our results with state-of-the-art MPS simulations on the cylinder, we find that the PEPS simulation provides very accurate results on the bulk properties \emph{and} the topological edge properties of the CSL.

%%%%%%%%%%%%%%%%%%
\par\noindent\emph{\textbf{Model and methods---}} %
%%%%%%%%%%%%%%%%%%
We investigate the model Hamiltonian,
\begin{multline}
    H = J_1 \sum_{\braket{ij}} \vS_i \cdot \vS_j + J_2 \sum_{\braket{\braket{ij}}} \vS_i \cdot \vS_j \\ + i\lambda \sum_{\braket{ijkl}_p} \left( P_{ijkl} - P_{ijkl}^{-1} \right),
\end{multline}
on an infinite square lattice, where $\vS_i$ are the spin-1/2 operators at site $i$, the first two sums are over nearest and next-nearest neighbors and the last sum is over all four-site plaquettes with $P$ the cyclic permutation operator. This model was constructed by taking a long-ranged parent Hamiltonian \cite{Nielsen2012} for the Kalmeyer-Laughlin (KL) state, and truncating the interaction terms \cite{Nielsen2013}; it was shown numerically that the ground state stays in the same KL phase for a large choice of parameters. Note that the last term explicitly breaks the time-reversal and reflection symmetries separately, such that a CSL ground state is stabilized. Here, we take the parameters $J_1=2\cos(0.06\pi)\cos(0.14\pi)$, $ J_2=2\cos(0.06\pi)\sin(0.14\pi)$ and $\lambda=2\sin(0.06\pi)$, approximately identified as the point for which the ground state has maximal overlap with the KL state \cite{Nielsen2013}. This choice of parameters was also considered in Ref.~\onlinecite{Poilblanc2017}, where a variational family of SU(2)-invariant PEPSs was used and the chiral nature of the ground state was confirmed.
\par A variational PEPS wave function can be represented as
\begin{equation}
    \ket{\Psi(A)} = \diagram{1},
\end{equation}
parametrized by a single tensor $A$ directly in the thermodynamic limit; the four virtual legs have bond dimension $D$ and are contracted, whereas the physical legs correspond to the spins in the lattice. We impose rotation and reflection-conjugation symmetry on the tensor
\begin{equation}
   \diagram{2} = \diagram{3} , \quad \diagram{4}=\diagram{3}.
\end{equation}
The tensor $A$ can be optimized by a variational energy minimization; here, we use gradient optimization \cite{Vanderstraeten2016}, where the energy is evaluated by CTMRG contraction and the energy gradient by automatic differentiation \cite{Liao2019,pepstorch}. It is the use of automatic differentiation that makes the unconstrained optimization possible, since its reverse mode evaluates gradient at only slightly higher cost, a multiplicative factor of $\mathcal{O}(1)$, than the energy itself irrespective of the number of variational parameters.
% Longer variant
%It is the use of automatic differentiation which makes the unconstrained optimization possible, since (i) its reverse mode evaluates gradient at only slightly higher cost than the energy itself, avoiding prohibitive scaling of finite differences, and (ii) even four-site plaquette interaction is elegantly accounted for without the need to introduce new diagrams as in the resummation methods~\cite{Corboz2016, Vanderstraeten2016}.
The CTMRG contraction induces a second control parameter, the environment bond dimension $\chi$. In our further calculations with the optimized PEPS, we use the CTMRG and variational uniform MPS (VUMPS) algorithms interchangeably for finding approximate environments, as these were recently shown to be equivalent \cite{Vanderstraeten2022}.
\par Additionally, we simulate this model on infinite cylinders with circumference $L_y$ using infinite MPS that are wound around the cylinder in a snakelike fashion. We have exploited SU(2) symmetry in the MPS simulation: because continuous symmetries necessarily remain unbroken on a finite-$L_y$ cylinder, imposing this symmetry on the MPS is not a restriction. We use the VUMPS algorithm to find an optimal MPS approximation for the ground state \cite{ZaunerStauber2018, Vanderstraeten2019a}. In our simulations, we go up to circumference $L_y=10$; with a truncation error of $\epsilon=10^{-5}$ we find that all reported observables have converged with a total bond dimension around $D_{\mathrm{MPS}}=15000$.

\begin{figure}
    \centering
    \includegraphics[width=\columnwidth]{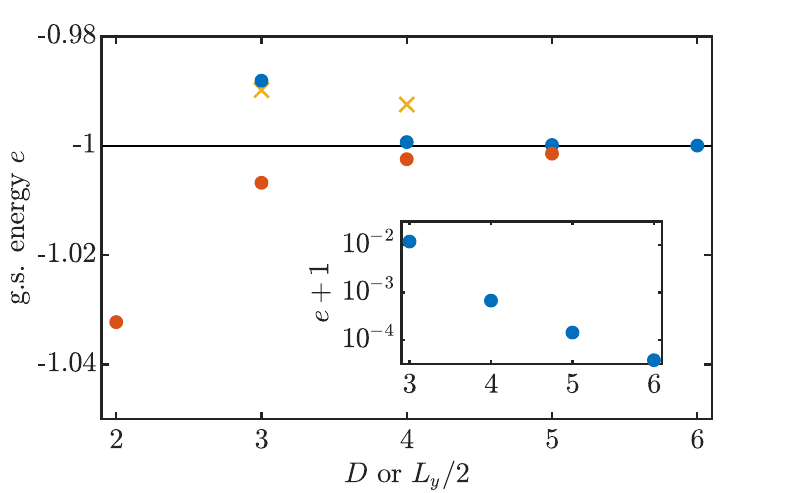}
    \caption{\textbf{Energetics.} Main: The ground state energy per site as a function of bond dimension $D$ for PEPSs (blue circles) and SU(2)-symmetric PEPS of Ref.~\cite{Poilblanc2017} (yellow crosses), or $L_y/2$ for cylinder-MPS (red circles). Inset: the variational PEPS energy compared to the value $e=-1$.}
    \label{fig:energetics}
\end{figure}

\begin{figure}
    \centering
    \includegraphics[width=\columnwidth]{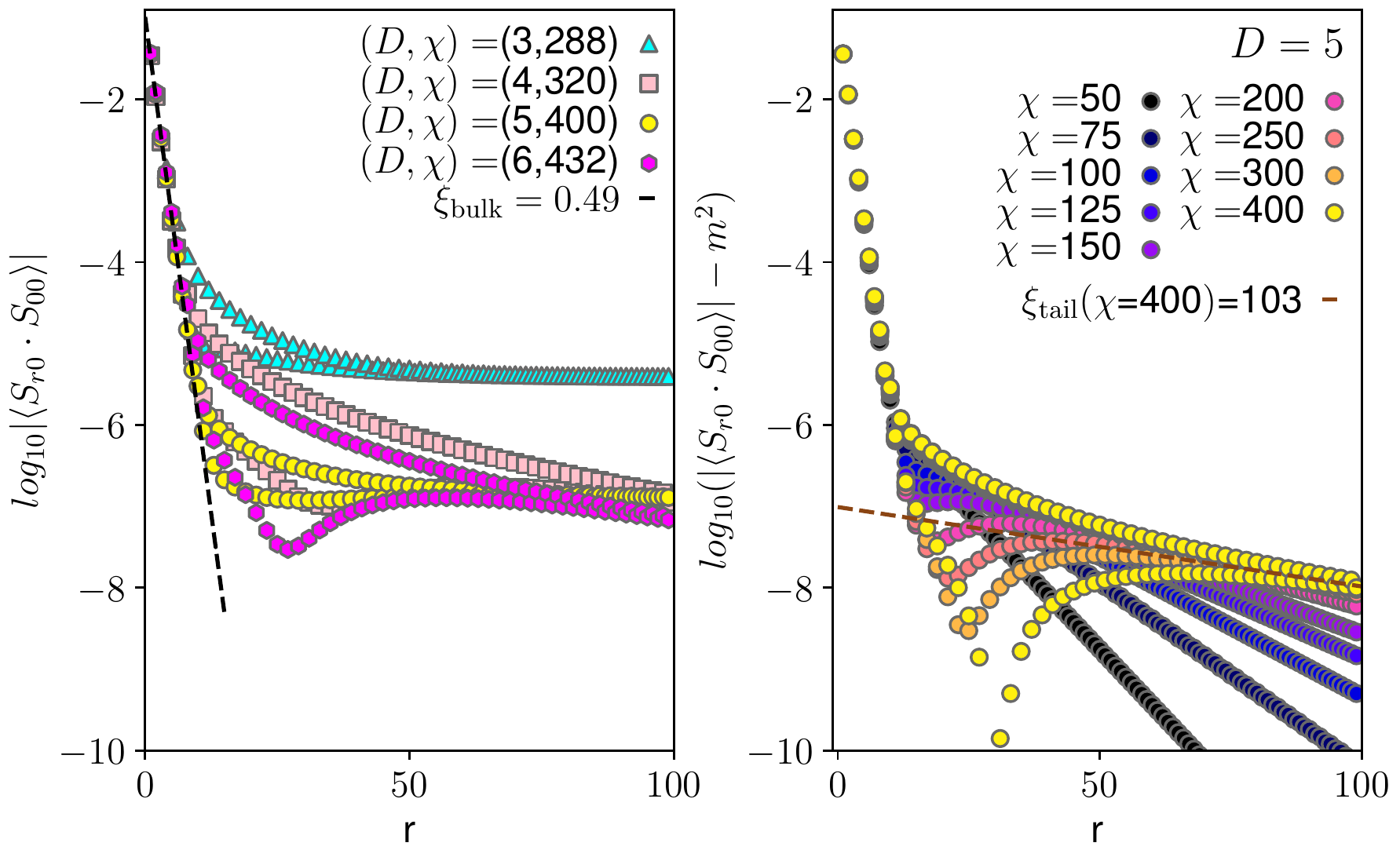}
    % \caption{\textbf{Long-range properties of PEPS CSL.} Left: Scaling of connected part of the real space spin-spin correlation functions $\langle \vec{S}(0)\cdot\vec{S}(r)\rangle_c$ for D=5 PEPS with environment dimension $\chi$. Right: Scaling of $\langle \vec{S}(0)\cdot\vec{S}(r)\rangle_c$ with bond dimension $D$.}
     \caption{\textbf{Long-range properties of PEPS CSL.} Left: scaling of the real space spin-spin correlation function $\langle \vec{S}_{r0}\cdot\vec{S}_{00}\rangle$ with bond dimension $D$
     at largest considered environment dimension $\chi$. Right: 
     scaling of connected part of the real space spin-spin correlation functions $\langle \vec{S}_{r0}\cdot\vec{S}_{00}\rangle-m^2$ for $D=5$ PEPS with $\chi$. For $D=5$ PEPS $m^2=1\times10^{-7}$.}
    \label{fig:correlations}
\end{figure}
%Scaling of correlation length by subleading gap $\delta_{14}=-log|\lambda_4/\lambda_1|$ in the transfer matrix $E$. The inset shows the leading part of the spectrum of $E$ as a function of the environment dimension $\chi$.

\begin{figure}
    \centering
    \includegraphics[width=\columnwidth]{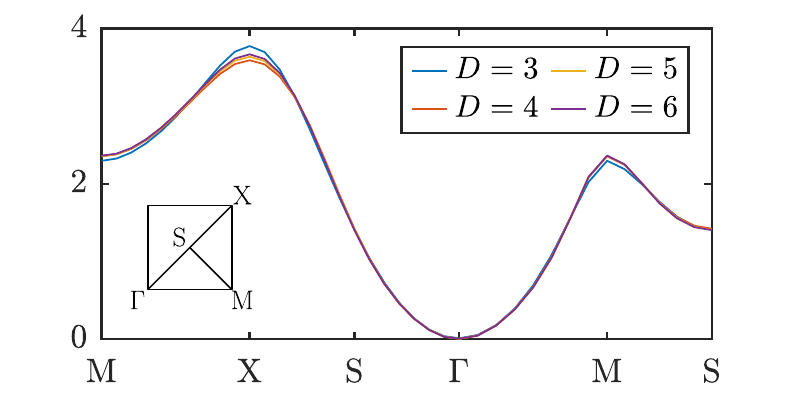}
    \caption{\textbf{Static structure factor.} The momentum-resolved correlation function [Eq.~\eqref{eq:sf}] for the $D=5$ PEPS for a path through the Brillouin zone.}
    \label{fig:sf}
\end{figure}

%%%%%%%%%%%%%%%%%%
\par\noindent\emph{\textbf{Bulk properties---}} %
%%%%%%%%%%%%%%%%%%
Let us first look at the variational energies that are obtained with both methods. In Fig.~\ref{fig:energetics} we plot the ground state energies. For $D=4$, we find a significantly better value of the energy than the SU(2)-invariant PEPS with the same bond dimension \cite{Poilblanc2017}. As the inset shows, the variational energy converges exponentially with $D$, which is in stark contrast with PEPS approximations for critical models, where a power-law extrapolation with the correlation length is typically needed \cite{Rader2018, Corboz2018}. We have also plotted the energy density as obtained from the (converged) cylinder-MPS simulations, where the finite circumference leads to an undershooting of the energy. From this comparison, it is clear that PEPSs give more accurate values for the ground state energy than the cylinder-MPS approach -- certainly given the fact that the largest MPS in our simulations contains around 70 million variational parameters [incorporating the SU(2) symmetry], compared to only around 2600 for the $D=6$ PEPS.
\par Surprisingly, we find that the energy converges to the value of $e=-1$. Given that the parameters in the model were not fine-tuned or chosen in a natural way, this feature is really peculiar and we did not find any satisfying explanation for this observation.
\par Let us then look at the magnetization given by $m^2=|\langle\vec{S}\rangle|^2$, which is equivalent to the spin-spin correlation function in the long-distance limit $m^2=\lim_{r\to\infty}\langle \vec{S}_{r0}\cdot\vec{S}_{00}\rangle$. In PEPS simulations of models at (or close to) criticality, order parameters are typically overestimated due to finite-$D$ effects, and a proper scaling of the order parameter is needed to get accurate results \cite{Rader2018, Corboz2018, Hasik2021}. Here, however, we find surprisingly small values of the magnetization already for small $D$: for $D=3$ we find a magnetization of $m^2=4\times10^{-6}$, which further decreases down to $m^2=6\times10^{-9}$ at $D=6$. This shows that the SU(2) symmetry of the CSL is realized in our PEPS description with extremely high accuracy.
% PREVIOUS VERSION
%\par Next, in Fig.~\ref{fig:correlations} we have plotted the structure of the correlations in the PEPS. The spin-spin correlation function along a horizontal row in the lattice is determined by the spectrum of the transfer matrix $E$ \cite{NISHINO199669,rams2018}, 
% \begin{equation}
%   E = \tmdiags{5}{0.15},
% \end{equation}
% where $a$ is a double-layer tensor with bond dimension $D^2$ and $T$ defines the boundary MPS approximation
% of the leading eigenvector of the PEPS transfer matrix 
% \begin{equation} \label{eq:peps-tm-eig-problem}
%   \tmdiags{1}{0.15} = \tmdiags{2}{0.15}, \quad \tmdiags{3}{0.15} \propto \tmdiags{4}{0.15}.
% \end{equation}
%We plot the leading eigenvalues of $E$ in Fig.~\ref{fig:correlations} (right panel) for the $D=5$ PEPS and increasing values of $\chi$. We observe that the gap in the transfer matrix approaches zero, indicating that the correlation length diverges. This is confirmed by the scaling of the correlation length to the infinite-$\chi$ limit; we find a correlation length of, at least, 450 sites. When plotting the spin-spin correlation function, however, we find an extended regime of exponential decay with a correlation length $\xi_{\mathrm{bulk}}=0.49$, and an algebraic decay with a magnitude of order $10^{-7}$. This indicates that the part of the transfer matrix with diverging correlation length gives only a very small contribution to the spin-spin correlations. For comparison we perform a similar analysis on a $D=3$ SU(2)-invariant PEPS in the Appendix.
%~\ref{appendix:d3-su2-peps-correlations}. 
% END PREVIOUS VERSION
\par Next, in Fig.~\ref{fig:correlations} we have plotted the structure of the correlations in the PEPS. The spin-spin correlation functions along a horizontal row in the lattice (left panel) show an extended regime of exponential decay with a short correlation length $\xi_{\mathrm{bulk}}=0.49$, consistent with a bulk gap. With increasing bond dimension $D$ this regime extends in range and the long-range order $m^2$ sharply decreases. However, for all $D$ considered, after the initial bulk-gapped regime the tails of the correlation functions decay in a very slow fashion. The data for the connected spin-spin correlations of the $D=5$ PEPS (right panel) show these tails
attaining large correlation lengths: for the largest $\chi$ considered $\xi_{\mathrm{tail}}(\chi=400)=103$ and an 
estimate based on scaling of the finite-$\chi$ data suggests $\xi_{\mathrm{tail}}(\chi\to\infty)\approx450$ at least. 
In a true CSL there are no such features. Here, as we argue below, these tails are an \textit{artefact} induced by the PEPS representation of a CSL. Nevertheless, their magnitude, $O(10^{-7})$ as indicated by the intercept with the $r=0$ axis, makes their contribution to spin-spin correlations minuscule. For comparison, we perform similar  analysis on both SU(2)-invariant and unconstrained $D=3$ PEPS in the Appendix.
\par Finally, in Fig.~\ref{fig:sf} we examine the momentum-resolved structure factor
\begin{equation} \label{eq:sf}
    s(q_x,q_y) = \sum_{mn} \e^{i(q_xm+q_yn)} \braket{\vS_{mn} \cdot \vS_{00} },
\end{equation}
which we evaluate using the methods from Ref.~\onlinecite{Vanderstraeten2022}. We plot this quantity through a path of the Brillouin zone for different PEPS bond dimensions, showing a smooth behavior at the $X$ point and a vanishing value at the $\Gamma$ point -- exactly what one would expect for a gapped spin liquid (long-range order would manifest itself as a divergence at the $X$ point). In addition, the curve has nicely converged as a function of PEPS bond dimension $D$. Note that this is the first instance where a spin liquid for a spin-$1/2$ system on a square lattice was found numerically by single-site PEPS, without imposing the SU(2) symmetry explicitly.

%%%%%%%%%%%%%%%%%%
\par\noindent\emph{\textbf{Entanglement spectrum---}} %
%%%%%%%%%%%%%%%%%%
As one of the defining characteristics of a chiral topological phase, we now investigate the entanglement spectrum of our variational PEPS ground states. Following the PEPS bulk-boundary correspondence \cite{Cirac2011}, the entanglement Hamiltonian $H_{\mathrm{ent}}$ can be inferred directly from the fixed point $\rho$ of the PEPS transfer matrix as $\rho=\e^{-\He}$. The fixed-point equation for PEPS transfer matrix, defined by a column (or row) of double-layer tensors $a$ with bond dimension $D^2$, reads
\begin{equation} \label{eq:peps-tm-eig-problem}
  \tmdiags{1}{0.15} = \tmdiags{2}{0.15}, \quad \tmdiags{3}{0.15} \propto \tmdiags{4}{0.15},
\end{equation} 
where the tensor $T$ defines the boundary MPS approximation with bond dimension $\chi$
 of the leading eigenvector of the PEPS transfer matrix. Numerically, we take the boundary MPS, fold it open and interpret it as a matrix-product operator (MPO) approximation for $\rho$. We have found above that $\rho$ has infinite correlation length in the infinite-$\chi$ limit, suggesting that $H_{\mathrm{ent}}$ has long-range interaction terms. Indeed, a thermal density operator for a local or exponentially decaying Hamiltonian could be efficiently represented as a MPO with finite bond dimension \cite{Molnar2015}.
\par
\begin{figure}
    \centering
    \includegraphics[width=\columnwidth]{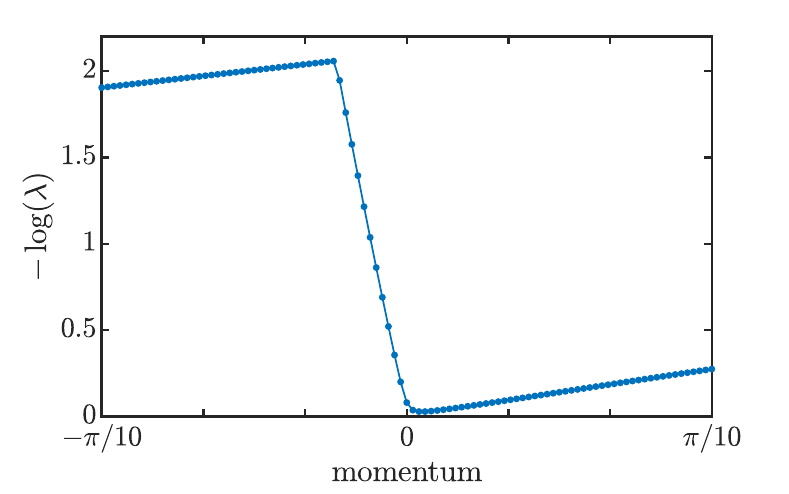}
    \includegraphics[width=0.49\columnwidth]{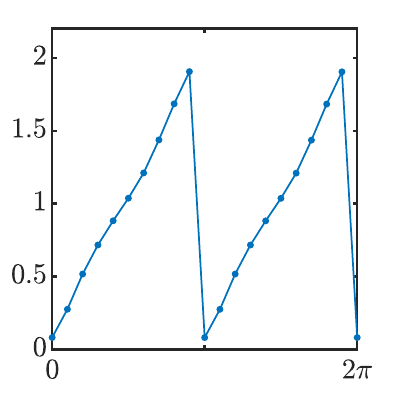}
    \includegraphics[width=0.49\columnwidth]{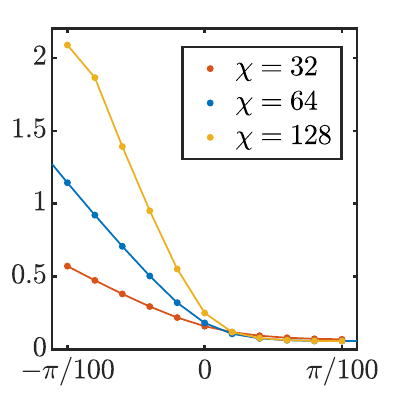}
    \caption{\textbf{Entanglement spectrum in thermodynamic limit.} We show results for the entanglement spectrum of the $D=5$ PEPS, using the quasiparticle ansatz. Top: a close-up for momenta around the origin, evaluated with $\chi=64$ and $\chi'=128$, nicely showing the two branches. Left: the dispersion in the full Brillouin zone. Right:  an even narrower close-up around the origin for three different values of $\chi$ and $\chi'=128$.}
    \label{fig:es_inf}
\end{figure}

 First, we compute the low-lying spectrum of $\He$ for the infinite system: we first find the leading eigenvector of $\rho$ as an infinite MPS with the VUMPS algorithm, and then apply a variational excitation ansatz with fixed momentum to capture the excited states of $\rho$ \cite{Haegeman2017, Vanderstraeten2019a}. A third control parameter enters here, the bond dimension $\chi'$ of this infinite MPS. In Fig.~\ref{fig:es_inf} we plot the results of this procedure for the $D=5$ PEPS. The lower-left panel shows the typical sawtooth form of a chiral spectrum in the full Brillouin zone, but the enlarged spectrum in the top panel around the origin shows that we also find a very steep mode. As the bottom-right panel shows, this additional mode gets steeper as the MPO's bond dimension $\chi$ is increased. This finding agrees with the fact that at finite $\chi$ the MPO has finite correlation length, and that $\He$ is not truly long-range; therefore, the spectrum cannot be perfectly chiral \cite{Nielsen1981} at finite $\chi$. Only in the infinite-$\chi$ limit can  we expect to find a perfectly chiral entanglement spectrum.
\par
\begin{figure}
    \centering
    \includegraphics[width=0.49\columnwidth]{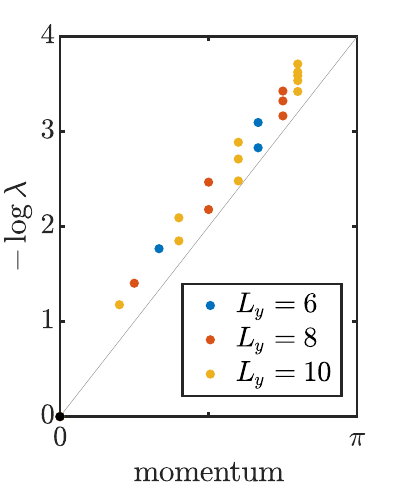}
    \includegraphics[width=0.49\columnwidth]{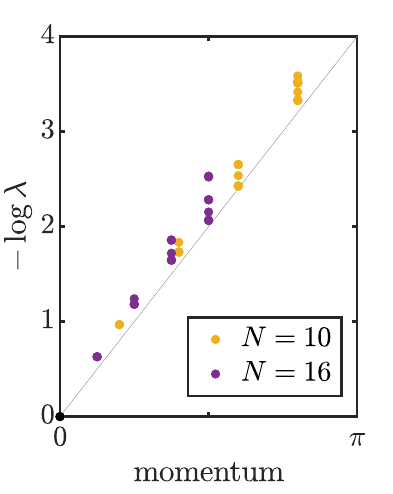}
    \caption{\textbf{Entanglement spectrum for finite system.} Left:  cylinder-MPS results for three different $L_y$. Right: spectrum of the MPO representation for the entanglement Hamiltonian $\rho$ confined on a ring of two different sizes $N$, as computed with the method in Ref.~\onlinecite{VanDamme2021} with $\chi=32$ and $\chi'=80$. Here we show the four first values of the momentum. We do not show the degeneracies of the SU(2) multiplets (for the cylinder-MPS these are exact; for the PEPS we find degeneracies up to the third or fourth digit).}
    \label{fig:es_fin}
\end{figure}

 Second, we compute the entanglement spectrum on a finite system of $N$ sites with periodic boundary conditions by placing the MPO approximation for $\rho$ on a finite ring. Since computing the spectrum exactly scales as $\mathcal{O}(D^N)$, it becomes quickly unfeasible for large $D$ or $N$. As an alternative, we use open-boundary MPS methods for finding the low-lying spectrum of MPOs on a ring \cite{VanDamme2021}; here, again, the bond dimension $\chi'$ enters as a control parameter. In Fig.~\ref{fig:es_fin} we show the entanglement spectrum of the $D=5$ PEPS confined on rings of different sizes, clearly showing the chiral branch with the degeneracies that are expected for the $\mathrm{SU}(2)_1$ Wess-Zumino-Witten conformal field theory. Moreover, we find the degeneracies of the SU(2) multiplets, again indicating that the PEPS is an SU(2)-invariant state up to high precision. In the left panel, we compare the spectra with the entanglement spectra as obtained from the MPS ground states on the cylinder; here, we use the usual prescription \cite{Cincio2013} for assigning momentum quantum numbers to the MPS entanglement spectrum. The two figures agree nicely, but the PEPS methodology allows us to reach much larger system sizes with moderate computational cost.

%%%%%%%%%%%%%%%%%%
\par\noindent\emph{\textbf{Conclusions---}} %
%%%%%%%%%%%%%%%%%%
This work shows that CSLs exhibiting topological order can be represented faithfully in terms of generic bosonic PEPSs upon variational optimization. The suspicion of an obstruction~\cite{Dubail2015} stems directly from the PEPS bulk-edge correspondence: the existence of a perfectly gapless chiral edge mode implies that the (one-dimensional) entanglement Hamiltonian~\cite{Cirac2011,Poilblanc2012} is long-range which, in turn, implies an infinite correlation length in the bulk. Nevertheless, our work shows that this feature does not prohibit us from variationally optimizing these PEPSs efficiently, and that the divergence of the correlation length manifests in a form of an artefact: an irrelevant vanishing long-range tail in bulk correlation functions. The rapid exponential convergence of the energy (much faster than the convergence of cylinder-MPS with circumference) and the vanishing of the magnetization are consistent with the expected bulk gap. Similarly, the expected topological properties in form of the chiral edge spectrum are obtained with high accuracy. The precise mechanism that gives rise to long-range tail artefact in PEPS description of CSL and its systematic dependence on bond dimension $D$ needs further investigation.
%\textcolor{red}{The systematic analysis of the weight of the long-range tails and and location of the crossover between short-range and long-range regime with bond dimension.}
\par Although the case of an SU($2$) Abelian CSL is considered here, we believe that chiral PEPSs can more broadly faithfully represent Abelian and non-Abelian SU($N$) CSL, $N\ge 2$, as suggested by calculations involving spin-symmetric PEPS~\cite{Chen2018,Chen2021}. We can also tackle cases of bosonic fractional Chern insulators such as the Harper-Hofstadter model \cite{Sorensen2005}, which can be realized in atomic simulators \cite{Aidelsburger2013, Miyake2013}. Finally, our results indicate that fermionic PEPSs \cite{Kraus2010, Corboz2010b} would be extremely well suited to describe electronic systems with chiral topological order like interacting Chern insulators at fractional filling~\cite{Bergholtz2013,Parameswaran2013} or chiral spin liquids close to a metal-insulator transition~\cite{Szasz2020, BBChen2021, Tocchio2021}.
\par Finally, going beyond ground-state properties, our work opens up the prospect of simulating bulk and edge excitation spectra \cite{Vanderstraeten2019b}, real-time dynamics, or finite-temperature properties \cite{Czarnik2019b} of chiral topological phases with PEPSs.

%%%%%%%%%%%%%%%%%%
\par\noindent\emph{\textbf{Acknowledgments---}} %
%%%%%%%%%%%%%%%%%%
We acknowledge inspiring discussions with Norbert Schuch and Frank Verstraete. We thank the Centro de Ciencias de Benasque Pedro Pascual, where this work was initiated. This work was supported by the Research Foundation Flanders, and the European Research Council (ERC) under the European Union's 
Horizon 2020 research and innovation programme (grant agreement no. 101001604). D. P. acknowledges support by the TNTOP ANR-18-CE30-0026-01 grant awarded by the French Research Council. This work was granted access to the HPC resources of CALMIP center under the allocation 2017-P1231.

\bibliography{./bibliography}

\clearpage
\appendix

%\onecolumngrid

\section*{Appendix: Long-range data for D=3 SU(2)-invariant PEPS}
\label{appendix:d3-su2-peps-correlations}
\begin{figure}[htbp]
    \centering
    \includegraphics[width=\columnwidth]{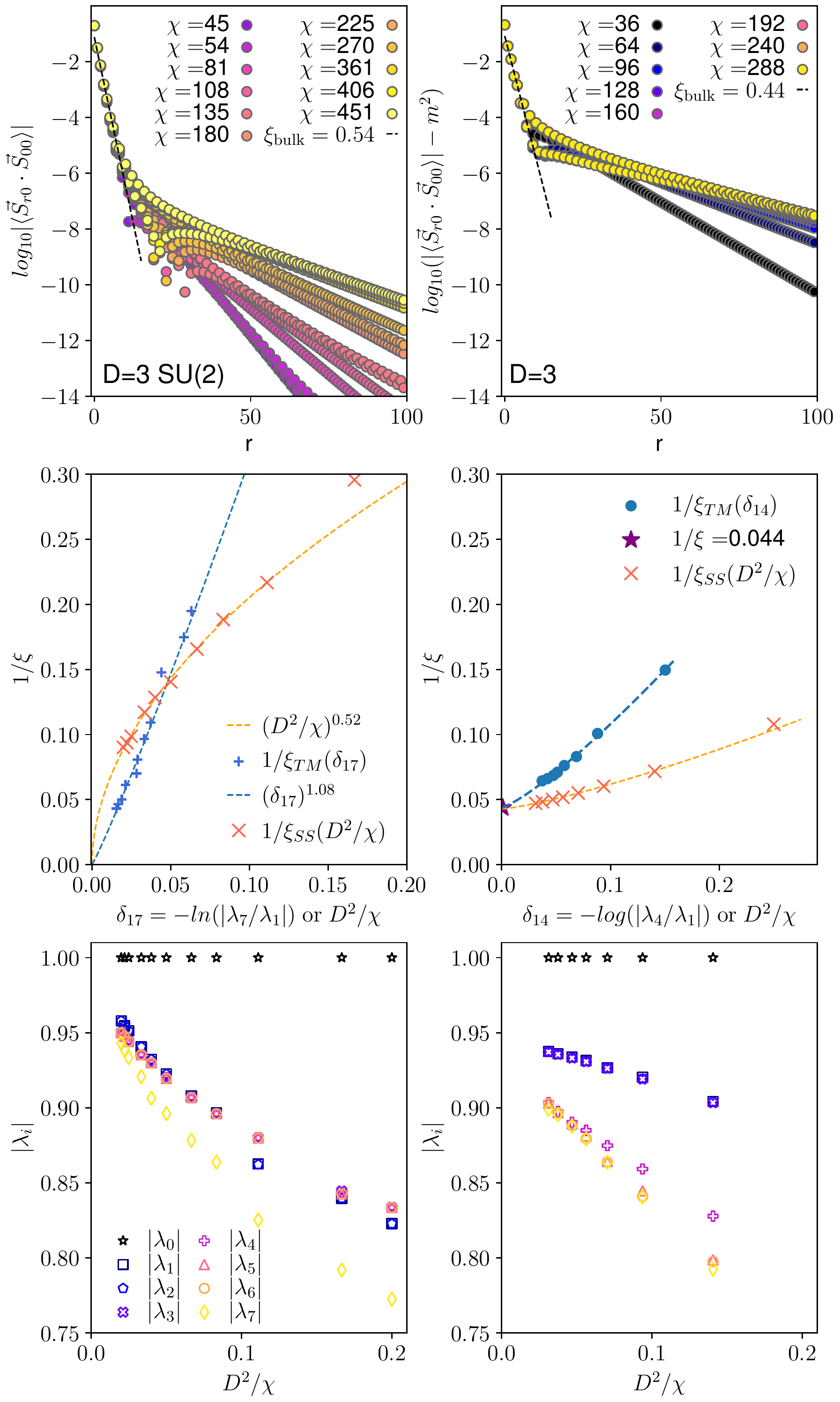}
    \caption{\textbf{Comparison of D=3 SU(2)-symmetric (left) and unconstrained PEPS (right).}   
    Top: Connected part of the real space spin-spin correlation functions. Unconstrained D=3 PEPS retain finite $m^2 = 4\times10^{-6}$.
    Center: Scaling of the correlation length by subleading gap in the transfer matrix $E$. For SU(2)-symmetric PEPS subleading gap is $\delta_{17}=-log|\lambda_7/\lambda_1|$. For unconstrained PEPS it is $\delta_{14}=-log|\lambda_4/\lambda_1|$ instead. Bottom: Leading part of the transfer matrix spectrum as a function of the environment dimension $\chi$.
    }
    \label{fig:app-su2-d3-correlations}
\end{figure}
The structure of spin-spin correlations $\langle \vec{S}_{r0}\cdot \vec{S}_{00}\rangle$ of PEPS,
represented by 2D tensor network with spin operators $\vec{S}$ inserted into two double-layer tensors
at distance $r$, can be analyzed by reduction to 1D tensor network. We make use of fixed-point of PEPS transfer matrix of Eq.~\ref{eq:peps-tm-eig-problem} to obtain a network analogous to 3-leg ladder
\begin{equation}
   \vcenter{\hbox{\includegraphics[scale=0.16]{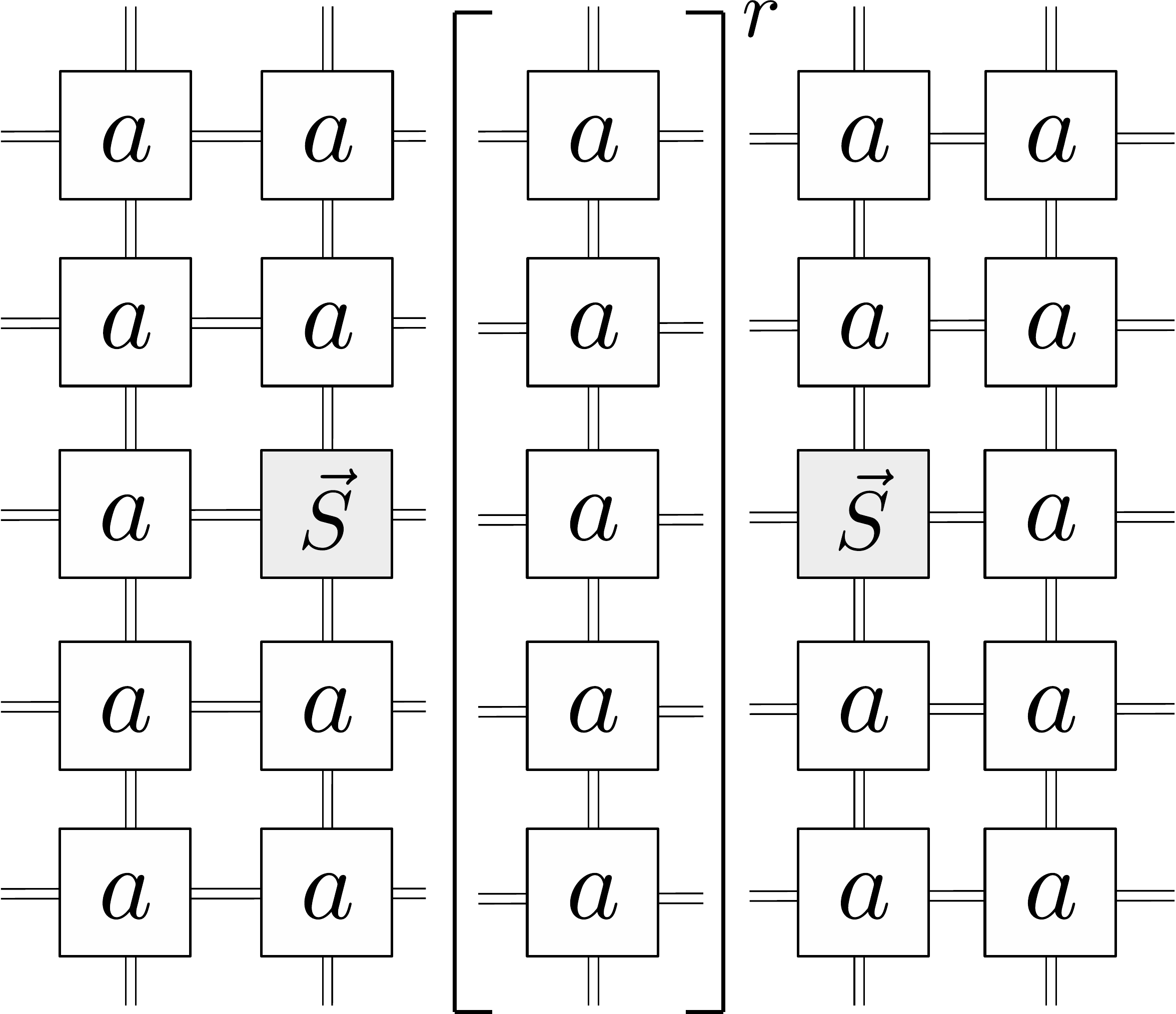}}} \approx \vcenter{\hbox{\includegraphics[scale=0.16]{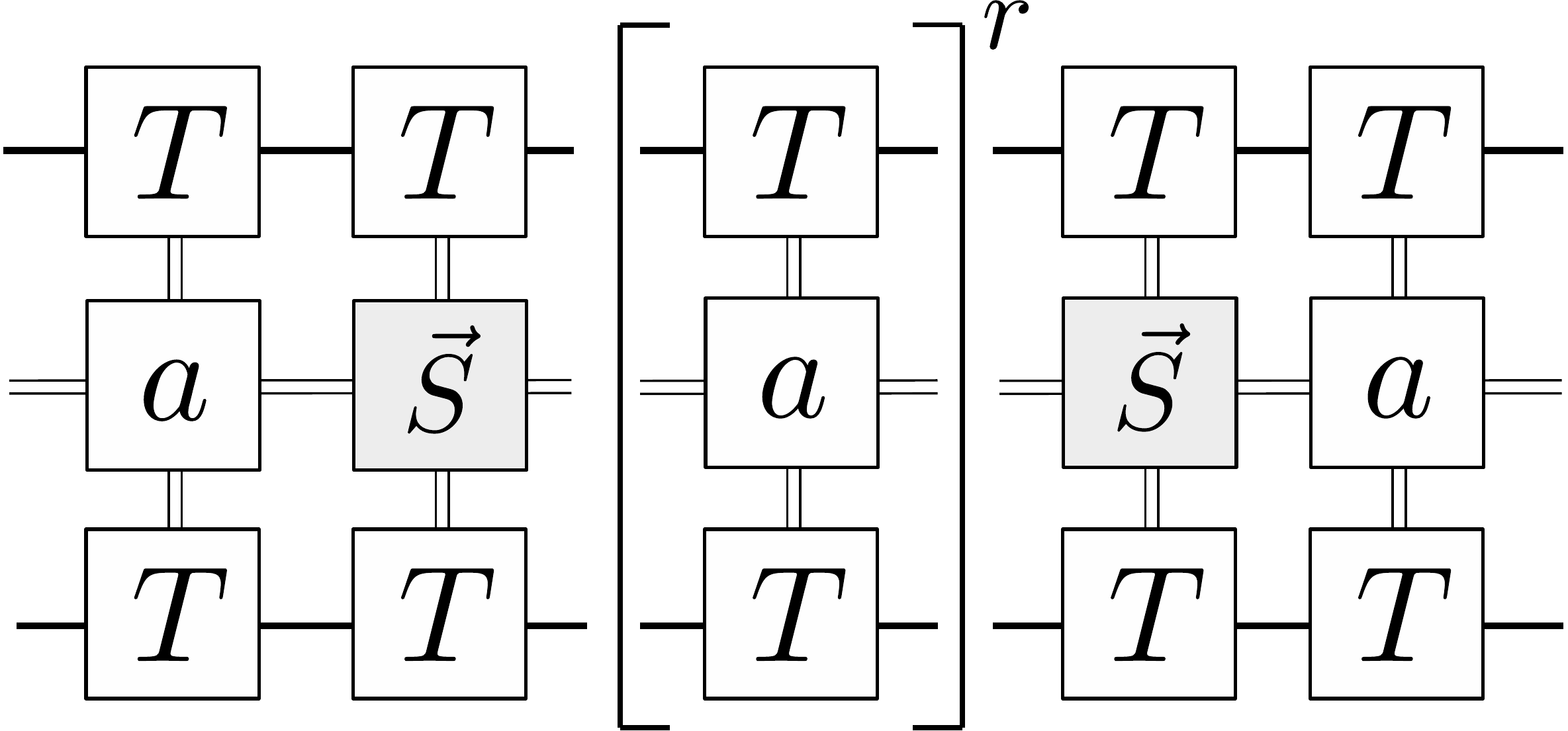}}}.
\end{equation}
The central $r$ rungs of this ladder define $\chi D^2\times\chi D^2$ transfer matrix $E$ raised to $r$-th power. The spectrum $\lambda_i$ of transfer matrix $E$ resolved as $E=\sum_i \lambda_i |r_i\rangle\langle l_i|$ then determines the decay of the spin-spin correlations, while the leading eigenvectors $|r_0\rangle, \langle l_0|$ define right and left boundary of this 3-leg ladder network.

\par In this appendix we present the scaling analysis of data for D=3 SU(2)-invariant PEPS ansatz, employed in \cite{Poilblanc2017} and compare it with the result for unconstrained D=3 PEPS.
The real-space spin-spin correlations $\langle S_{r0}\cdot\vec{S}_{00}\rangle$, scaling of the correlation length $\xi$, and the leading part of the spectrum of transfer matrix $E$ are presented in Fig.~\ref{fig:app-su2-d3-correlations}. The real-space spin-spin correlations show two distinct regimes analogous to the unrestricted
PEPS ansatz: Up to a range of roughly 10 sites a fast exponential decay, with $\xi_{\mathrm{bulk}}=0.54$ for D=3 SU(2) ansatz, then crossing over into regime of extremely slow decay with strong finite-$\chi$ dependence.
Unlike unconstrained PEPS with finite $\xi \approx 23$, the tail of the SU(2)-symmetric ansatz shows genuine
algebraic decay in the $\chi\to\infty$ limit. This feature is revealed by the  analysis of the spectrum of the
transfer matrix $E$ and decay of the tail of spin-spin correlations.
In particular, we scale finite-$\chi$ data in two different ways: First, by extracting the spin-spin correlation length $\xi_{SS}$ from the tails of spin-spin correlations 
and performing a power-law fit with respect to the environment dimension $\chi$. Second, by scaling the correlation length $1/\xi_{TM}$ given by the leading gap in the transfer matrix $E$ against the subleading gap and again performing power-law fit as suggested by Ref.~\cite{rams2018}. 
%Note, that unlike in the case of unrestricted D=3 PEPS the order of the eigenvalues of $E$ is not fixed. For large $\chi$, the subleading eigenvalue $\lambda_1$ is two-fold degenerate but for $D^2\chi \lesssim 0.1$ the subleading eigenvalue $\lambda_3$ is instead four-fold degenerate. 
Note that for SU(2)-symmetric PEPS the data suggest $\lambda_7$ to define subleading gap for larger $D^2/\chi$ and we thus use it for the purpose of the scaling.

\end{document}